\pdfoutput=1
\documentclass[11pt,a4paper]{article}
\usepackage{graphicx}
\usepackage{amsmath}
\usepackage{amssymb}
\usepackage{times}
\usepackage{verbatim}
\DeclareGraphicsExtensions{.pdf}
\usepackage{amssymb}

\begin{document}

\title{Evolution of Coordination in  Social Networks: A Numerical Study}
\author{\vspace{.2cm}
M. Tomassini\footnote{corresponding author: marco.tomassini@unil.ch} \hspace{1.5cm}
E. Pestelacci\\
\small{Information Systems Department, HEC}\\
\small{University of Lausanne,Switzerland}
}
\maketitle

\begin{abstract}
Coordination games are important  to explain efficient and desirable social behavior. Here we study these games
 by extensive numerical simulation on networked social structures using an evolutionary approach.
 We show that local network effects may promote selection of efficient equilibria in both pure
and general coordination games and may explain social polarization. These results are put into perspective with 
respect to known theoretical results. The main
insight we obtain is that clustering, and especially community structure in social networks has a positive role in
promoting socially efficient outcomes.

\noindent

\end{abstract}

\section{Introduction}
\label{intro}

Game theory~\cite{vega-redondo-03} has proved extremely useful in the study
of economic, social, and biological situations
for describing interactions between agents having possibly different and often conflicting objectives.
Paradigmatic games such as the Prisoner's Dilemma~\cite{axe84}  have been used 
in order to represent the tension that appears in society when
individual objectives are in conflict with socially desirable outcomes. 
Most of the vast research literature has focused on conflicting situations in order to
uncover the mechanisms that could lead to cooperation instead of socially harmful 
outcomes (see e.g.~\cite{nowak-06} for a synthesis).
However, there are important situations in society that do not require players to use
aggressive strategies. In fact, many frequent social and economic activities require individuals to
coordinate their actions on a common goal since in many cases the best course of action
is to conform to the standard behavior. For example, if one is used to drive on the right side of the road 
and travels to a country where the norm is reversed,
it pays off to follow the local norm. Bargaining  and contracts are also of this type because, even though
expectancies may be different between a buyer and a seller, still both would rather trade than
not, provided that the respective prices are not too different. 
For another example, consider a situation in which coordination in working contributions is required in order to produce
a good or a service. In a group it might pay off not to contribute, if this behavior goes unnoticed, but the total output will 
be negatively affected.
Games that express this
extremely common kind of interactions are called \textit{coordination games}. 

Coordination games confront the players with multiple Nash equilibria and the
ensuing problem of equilibrium selection. Given that these equilibria are equivalent from
the game-theoretical point of view,
how to explain how agents make their decisions? A useful approach has been to use evolutionary
and learning ideas  which offer a dynamical perspective based on the forces of biological and social
evolution. In \textit{evolutionary game theory} (EGT), the concept
of a population of players where strategies that score best are more likely to be selected and reproduced 
provides a justification for the appearance of stable states of the dynamics that represent solutions
of the game~\cite{vega-redondo-03,weibull95}.\\
For mathematical convenience, standard EGT is based on infinite mixing populations 
 where pairs of individuals are drawn uniformly at random at each
step and play the game. Correlations are absent by definition and the population has an
homogeneous structure. However, everyday observation tells us that in animal and human
societies, individuals usually tend to interact more often with some specified subset of
partners; for instance, teenagers tend to adopt the fashions of their close friends group; closely
connected groups usually follow the same religion, and so on. Likewise, in the economic world,
a group of  firms might be directly connected because they share capital, technology, or
otherwise interact in some way.
In short, social interaction is mediated by networks, in which vertices identify people,
firms etc., and edges identify some kind of relation between the concerned vertices such as
friendship, collaboration, economic exchange and so on. Thus, locality of interaction plays
an important role. This kind of approach was pioneered in EGT by Nowak and May~\cite{nowakmay92} 
by using simple two-dimensional regular grids. Recently, in the wake of  a surge
of activity in network research in many fields~\cite{newman-03,social-nets-05}, the dynamical and evolutionary
behavior of games on networks that are more likely to represent actual social interactions than regular grids 
has been investigated (see~\cite{Szabo-Fath-07} for a comprehensive recent review). These studies, almost exclusively conducted on
games of conflict such as the Prisoner's dilemma or Hawks-Doves, have shown that
there are network structures, such as scale-free and actual social networks that may favor the emergence of 
cooperation with respect to 
 the fully mixing populations used in the 
theory~\cite{santos-pach-06,luthi-pest-tom-physa08}. 

In this work we extend this kind of approach to games of the coordination type. We shall use several
types of network structures, both networks generated by an algorithm as well as an actual social network
to try to unravel the effect of structure on the population behavior.
In the present paper, we ignore that social networks are actually dynamical entities that change constantly.
Indeed, actors join and leave networks and they may accumulate and abandon ties over time.
Using static networks is a useful first approximation however, especially for the cases where the rate of change
of the network structure is slow with respect to the rate of change of individual's behaviors which is the approximation
that is made here.\footnote{a companion study on
the dynamical network case is in progress.}
Comparatively little theoretical work has been done on coordination games on networks, except for some standard
types such as rings or complete networks~\cite{ellison} for which rigorous results have been obtained thanks to
their regular structure. Although we do mention some known rigorous results as discussed below, our methodology is essentially 
computer simulation-based. 
This is because for most network types,
inhomogeneity and correlations do not allow standard mean-field methods to be used. Likewise, pair approximation
methods~\cite{vbaalen00} provide an acceptable approach for random and regular graphs but not for the other more
complex types and thus they are not used here.\\
The paper is organized as follows. In the next section we first present a brief  introduction to the subject
of coordination games, in order to make the work self-contained. Then, in Sect.~\ref{maths}, we enumerate the main theoretical 
results on coordination games, as well as the necessary definitions for networks of agents and their dynamics.
In Sect.~\ref{num} we describe the simulation methodology and the parameters used and, in Sect.~\ref{sim-results} we present
and discuss the simulation results on various network classes first for pure coordination games, and then for general coordination ones. Finally, in Sect.~\ref{concl} we give our conclusions and ideas for future work.

\section{Coordination Games}
\label{coord-games}

\subsection{General Coordination Games}
\label{gen-coord}

General two-person, two strategies coordination games have the normal form of Table~\ref{coord-game}. 
With $a > d$ and $b>c$,
~$(\alpha,\alpha)$ and $(\beta,\beta)$ are both Nash equilibria. Now, if we assume that $a>b$ and $(a-d) \le (b-c)$
then $(\beta,\beta)$ is the risk-dominant equilibrium, while $(\alpha,\alpha)$ is the Pareto-dominant one.
This simply means that players get a higher payoff by coordinating on $(\alpha,\alpha)$ but they risk less
by using strategy $\beta$ instead. There is also a third equilibrium in mixed strategies but it is evolutionarily unstable.
\begin{table}[!ht]
\begin{center}
\begin{tabular}{c|cc}
 & $\alpha$ & $\beta$\\
\hline
$\alpha$ & $a,a$ & $c,d$\\
$\beta$ & $d,c$ & $b,b$\\
\end{tabular}
\caption{A general two-person, two strategies coordination game.\label{coord-game}}
\end{center}
\end{table}
A well known example of games of this type are the so-called Stag-Hunt games~\cite{skyrms}.  This class of games has been extensively
studied analytically in an evolutionary setting~\cite{kandori,ellison} and by numerical simulation on several model network types~\cite{skyrms,santos-pach-06,luthi-pest-tom-physa08,anxo1}.
In the following, we shall first deal with the easier case of pure coordination games which, in spite of their simplicity,
already clearly pose the equilibrium selection problem. Then we shall report results on Stag-Hunt games, for which there exist
many published studies to compare with, both theoretical and with the use of simulation.

\subsection{Pure Coordination Games}
\label{pure-coord-games}

Two-person \textit{pure coordination games} have the 
normal form depicted in Table~\ref{coord}, with $u_i,u_i>0$, and $u_i,u_j =0,0, i \ne j,  \forall i,j \in [1,k]$, where $k$ is the number
of strategies available to each player in the strategy set $S=\{s_1,s_2,...,s_k\}$, and the $u$'s are payoffs. So all the Nash equilibria in pure strategies correspond to diagonal
elements in the table where the two players coordinate on the same strategy, while there is a common lower 
uniform payoff for all other strategy pairs which is set to $0$ here.
\begin{table}[!ht]
\begin{center}
\begin{tabular}{c|cccc}
 & $s_1$ & $s_2$ & $\ldots$ & $s_k$\\
\hline
$s_1$ & $\bf u_1,u_1$ & $0,0$ & $\ldots$ & $0,0$\\
$s_2$ & $0,0$ &  $\bf u_2,u_2$ & $\ldots$ & $0,0$\\
$\ldots$ & $\ldots$ & $\ldots$ & $\ldots$ & $\ldots$\\
$s_k$ & $0,0$ & $\ldots$ & $\ldots$ & $\bf u_k,u_k$
\end{tabular}
\end{center}
\vspace{-0.1cm}
\caption{A general payoff bi-matrix of a two-person pure coordination game. Nash equilibria in pure strategies are marked in bold.}
\label{coord}
\end{table}
A simple coordination game is the \textit{driving game}. In some countries people drive
on the right side of the road, while in others they drive on the left side. This can be represented
by the  pure coordination game represented in Table~\ref{drive}.
\begin{table}[!ht]
\begin{center}
\begin{tabular}{c|cc}
 & right & left\\
\hline
right & $\bf 1,\bf 1$ & $0,0$\\
left & $0,0$ & $\bf 1,\bf 1$\\
\end{tabular}
\caption{The driving game.\label{drive}}
\end{center}
\end{table}
\noindent There are two Nash equilibria in pure strategies: (right, right) and (left, left) and obviously
there is no reason, in principle, to prefer one over the other, i.e.~the two equilibria are equivalent. However, while some countries have got accustomed
to drive on the left such as the UK, Australia, and Japan, others have done the opposite such as
most European countries and the USA. 
Such \textit{norms} or \textit{conventions} have stabilized in
time and are often the product of social evolution. There is of course a third equilibrium
in mixed strategies in the driving game which consists in playing left  and right with probability $1/2$
each but it would seem rather risky to play the game in this way on a real road. Another well known example of a pure
coordination game is the Battle of the Sexes in which the Nash equilibria in pure strategies are those in which
players use the same strategy, but the two sides in a two person game prefer a different equilibrium~\cite{vega-redondo-03}.

\section{Mathematical Setting and Previous Results}
\label{maths}

In this section, we recall some rigorous results for two-person, two-strategies coordination games on some particular network
types. Indeed, network topology has an influence on the stable states of the
evolutionary dynamics that will be reached, as it will become clear in what follows. We also give nomenclature and definitions for the graphs
representing the population of agents and for the dynamical decision processes implemented by the agents. 

Let's thus consider the 
game's payoff matrix of Table~\ref{coord_game} with $a \ge b > 0$. When $a > b$, strategy $\alpha$ is said to be dominant
since a player obtains a higher payoff playing $\alpha$ rather than $\beta$.
\begin{table}[!ht]
\begin{center}
\begin{tabular}{c|cc}
 & $\alpha$ & $\beta$\\
\hline
$\alpha$ & $a,a$ & $0,0$\\
$\beta$ & $0,0$ & $b,b$\\
\end{tabular}
\caption{A general two-person, two-strategies pure coordination game.\label{coord_game}}
\end{center}
\end{table}

The network of agents will be represented by an undirected graph $G(V,E)$, where the
 set of vertices $V$ represents the agents, while the set of edges (or links) $E$ represents their symmetric interactions. The
 population size $N$ is the cardinality of $V$. A neighbor  of an agent $i$ is any other agent $j$ at distance one
 from $i$. The set of neighbors of $i$  is called  $V_i$  and its cardinality is the degree $k_i$ of vertex $i \in V$. The average
degree of the network is called $\bar k$ and $p(k)$ denotes its degree distribution function, i.e. the probability
that an arbitrarily chosen node has degree $k$. 

Since we shall adopt an evolutionary approach, we must next define the decision rule by which individuals will
update their strategy during time. An easy and well known adaptive learning rule is \textit{myopic best-response
dynamics}, which embodies a primitive form of bounded rationality and for which rigorous results are known~\cite{young,Goyal}. 
In the local version of this model, time
is discrete i.e. $t =0,1,2,\ldots$ and, at each time step, an agent has the opportunity of revising her current
strategy. She does so by considering the current actions of her neighbors and switching to the action that
would maximize her payoff if the neighbors would stick to their current choices. The model is thus completely
local and an agent only needs to know her own current strategy, the game payoff matrix, who are her neighbors, and
their current strategies. This rule is called myopic because the agents only care about immediate payoff,
they cannot see far into the future. Given the network structure of the population, the rule is implemented as
follows: 
\begin{itemize}
\item at each time step a player $i$ revises his strategy with probability $p$
\item player $i$ will choose the action that maximizes his payoff, given that the strategy
profile of his neighbors $V_i$ remains the same as in the previous period
\item if there is a tie or $i$ is not given the opportunity of revising his strategy, then
$i$ will keep his current strategy
\end{itemize}

Using the above kind of stochastic evolutionary process, which can be modeled by a Markov chain,
the following theoretical results have been proved by several researchers and can be found in Chapter $4$ of~\cite{Goyal}, 
where references to the original works are given. They are valid for general coordination games, and thus also for the
special case of the pure coordination game of Table~\ref{coord_game}.

\noindent {\bf Theorem.} A strategy profile in which everyone plays the same action is a Nash equilibrium for
every graph $G$.  If $G$ is complete then these are the only possible equilibria.  If $G$ is
incomplete, then there may exist polymorphic equilibria as well.

The preceding theorem implies that social diversity may emerge at equilibrium depending on the network structure. Given
that complete networks are not socially relevant, this result leaves open the possibility of equilibrium strategy distributions 
in the population. A second related result states that, starting from any initial strategy profile, the above described stochastic process 
will converge to a Nash equilibrium of the coordination game with probability $1$. To probe for the stability of equilibria, 
the concept of mutation
 is introduced. A mutation simply means
that a player that is updating its current strategy can make a mistake with some small probability $q$. 
These small random effects are meant to capture various sources of uncertainty such as deliberate and involuntary
decision errors. Deliberate errors might play the role of experimentation in the environment, and involuntary ones
might be linked with insufficient familiarity with the game, for example.
A state of this adaptive noisy dynamics is called \textit{stochastically stable} if in the long term, the probability of being in that
state does not go to tero as the error probability tends to zero
(see~\cite{young} for
a rigorous definition). This idea allows one to discriminate among  the possible equilibria according to their stability
properties. \\

From the above considerations, it may be concluded that the network topology plays an important role
on the equilibrium states that the population will reach in the long run. However, the graph types for which analytical
results are available are far from the complex structures of observed real social networks. Therefore, our aim
in the following is to characterize the behavior of such complex networks by using numerical simulations and
appropriate statistical analysis. 

\section{Numerical Simulations Methodology}
\label{num}

\subsection{Network Types Studied}
\label{net-types}
 
In the last few years a large amount of knowledge has accumulated  about
the structure of real social networks and many model networks, both static and growing have been 
proposed~\cite{newman-03,social-nets-05,jackson-book}.
We are thus in a position that allows us to make use of this recent information in order to study the behavior
of coordination games on such realistic networks. In detail, we shall use the following network types: random, 
Bar\'abasi-Albert scale-free networks, a real social network, and model social networks. We shall now briefly 
describe each of these network types, directing the reader to the
relevant references for more details.

\subsubsection{random graphs}

For generating random graphs we use one of the classical models proposed by Erd\"os and R\'enyi and described
 in~\cite{bollobas}. Given $N$ indistinguishable vertices, each possible edge
has an independent probability $p$ of appearing ($0 \le p \le 1$), which gives the $G(N,p)$ ensemble of random graphs. It is worth mentioning that for that type of random graph the average clustering 
coefficient\footnote{We use the following common definition. The clustering coefficient
$C_i$ of a node $i$ is defined as $ C_i=2E_i/k_i(k_i-1)$, where $E_i$ is the number of edges in  the
neighborhood of $i$. Thus $ C_i$ measures the amount of ``cliquishness'' of the
neighborhood of node $i$ and it characterizes the extent to which nodes adjacent to node $i$ are
connected to each other. The clustering coefficient of the graph is the average over all nodes:
$\bar C = \frac{1}{N} \sum_{i=1}^{N} C_i$~\cite{newman-03}}
$\bar C=p=\bar k/ N$. Thus $\bar C$ at fixed $\bar k$ tends to $0$ for increasing $N$. This is one of the reasons
that make these random graphs rather unsuitable as model social networks, although they are useful as
a known benchmark to evaluate deviations from randomness. Furthermore, $p(k)= e^{-\bar k}\; \frac{ \bar k^k}{k!}$
is Poissonian and thus it allows only small fluctuations around $\bar k$, while actual measured
networks usually have long-tailed degree distribution functions.

 \subsubsection{Scale-Free graphs} 
 
 Among the several available models for constructing scale-free networks~\cite{newman-03}, here we
 use the classical one by Ba\-ra\-b\'asi--Albert~\cite{alb-baraba-02}.
 Ba\-ra\-b\'asi--Albert networks are grown incrementally starting with a small clique 
of $m_0$ nodes.
At each successive time step a new node is added such that its $m \le m_0$ edges link it to
$m$ nodes already present in the graph. It is
assumed that the probability $p$ that a new node will be connected to node $i$ depends on
the current degree $k_i$ of the latter. This is called the \textit{preferential attachment} rule. 
The probability $p(k_i)$ of node $i$ to be chosen is given by $p(k_i) = {k_i}/ \sum_{j} k_j,$ 
where the sum is over all nodes already in the graph.
The model evolves into a stationary network
with power-law probability distribution for the vertex degree $P(k) \sim k^{-\gamma}$, with
$\gamma\sim 3$. 
For the simulations, we started
with a clique of $m_0=2$ nodes and at each time step the new incoming node has $m=2$ links.\\
Scale-free graphs are rather extreme and are  infrequent among social networks (see below), even taking
finite degree cutoffs into account. As the random graph, they are rather to be considered as a model
network.

\subsubsection{An Actual Social Network}

One important reason for introducing true or model social networks is that, as said above, clustering is an
important feature in networks of contacts while neither Erd\"os-R\'enyi nor Barab\'asi-Albert scale-free graphs
show a comparable level of clustering.
As a typical example of a true social network, we use a coauthorship network among researchers in the genetic programming (GP) community. This 
network has a connected giant component of 1024 scientists and it has recently been analyzed \cite{gp-graph-gpem}. It has clusters and communities and it should be representative
of other similar human acquaintance networks. Its degree distribution function $p(k)$, as is usually the case
with most measured social networks~\cite{am-scala-etc-2000,newman-03,jackson-book}, is not a pure power-law;
rather, it can be fitted by an exponentially truncated power-law.

\subsubsection{Model Social Networks}

Several ways have been proposed for growing artificial networks with properties similar to those of
observed social networks. Here we use the model of Toivonen et al.~\cite{toivonen-2006}, which
was conceived to construct a graph
with most of the desired features of real-life social networks i.e, assortative, highly clustered, 
showing community structures, having an adjustable decay rate
of the degree distribution, and a finite cutoff.
The network is incrementally grown starting from a seed of $m_0$ randomly connected vertices.
At each successive time step, the following algorithm is applied:
\begin{enumerate}
\item On average $m_r \geq 1$ random vertices are picked to be initial contacts.
\item \label{sec_contacts} On average $m_s \geq 0$ neighbors of the $m_r$ initial contacts are
chosen to be secondary contacts.
\item A newly added vertex $v$ is connected to all the initial and secondary contacts determined in the two previous steps.
\end{enumerate}
The above is iterated until the network reaches the desired size.
Notice that the process responsible for the appearance of high clustering, assortativity and community structure is step \ref{sec_contacts}.
In the numerical experiments, we used graphs of size $N=1000$ 
with $m_0 = 30$ initial nodes. Every time a new node is added, its
number of initial contacts $m_r$ is distributed as
$p(\#\textrm{ of initial contacts} = 1) = 0.95$ and $p(\#\textrm{ of initial contacts} = 2) = 0.05$.
The number of its secondary contacts $m_s$  is uniformly distributed between 0 and 3. The resulting degree
distribution falls below a power-law for high values of $k$~\cite{toivonen-2006}.

\subsection{Simulations Settings}
\label{params}

The network used are of size $N=1000$ except for the GP network, whose giant component has size 1024. 
The mean degree $\bar k$ of the networks generated was $6$, except for the GP case which has $\bar k \simeq 5.8$.\\
For pure 
coordination games the non-zero diagonal payoffs $a$  (see sect.~\ref{pure-coord-games}) has been varied in the range
$[0.5, 1]$ in steps of $0.05$ with $b=1-a$; the range $[0, 0.5]$ is symmetrically equivalent. For general coordination
 games (sect.~\ref{coord-games}) in which $a > d > b > c$, we have studied a portion of the parameters' space defined by $c \in [-1,0]$ and
 $d \in [0,1]$, $a=1$, and $b=0$,  as is usually done for the stag-hunt games~\cite{santos-pach-06,anxo1}. The $c-d$ plane has been sampled
 with a grid step of $0.05$.\\
 Each value in the phase space reported in the following figures is the average of $50$ independent runs. Each run has 
 been performed on a fresh
 realization of the corresponding graph, except for the GP co-authorship network case which is a unique realization.\\
 As already hinted  in sect.~\ref{maths}, we have used a fully asynchronous update scheme in which a randomly
 selected agent is chosen for update with replacement at each discrete time step. 
 To detect steady states\footnote{True equilibrium states in the sense of stochastic stability are not guaranteed
 to be reached by the simulated dynamics. For this reason we prefer to use the terms steady states or quasi-equilibrium states
 which are states that have little or no fluctuation over an extended period of time. } of the dynamics we first let the system evolve for a transient period of $5000 \times N \simeq 5 \times 10^6$ time steps.
 After a quasi-equilibrium state is reached past the transient, averages are calculated during $500 \times N$ additional
 time steps. A steady state has always been reached in all simulations performed within the prescribed amount of
 time, for most of them well before the limit. \\
 We have experimented with different proportions of uniformly randomly distributed initial strategies $\alpha$ belonging
 to the set $\{0, 0.05, 0.25, 0.5, 0.75, 0.95, 1\}$  and
 we have used two different values for the stochastic noise $q$ in the simulations: $q \in \{0, 0.02\}$, i.e. either
 no noise or a small amount, as prescribed by the most important the\-ore\-tical stochastic models in order to ensure
 that the evo\-lu\-tio\-nary process is er\-godic~\cite{kandori,ellison,young}.

\section{Simulation Results}
\label{sim-results}

\subsection{Results on Pure Coordination Games}

\begin{figure} [!ht]
\begin{center}
\begin{tabular}{cc}
\includegraphics[width=6.5cm] {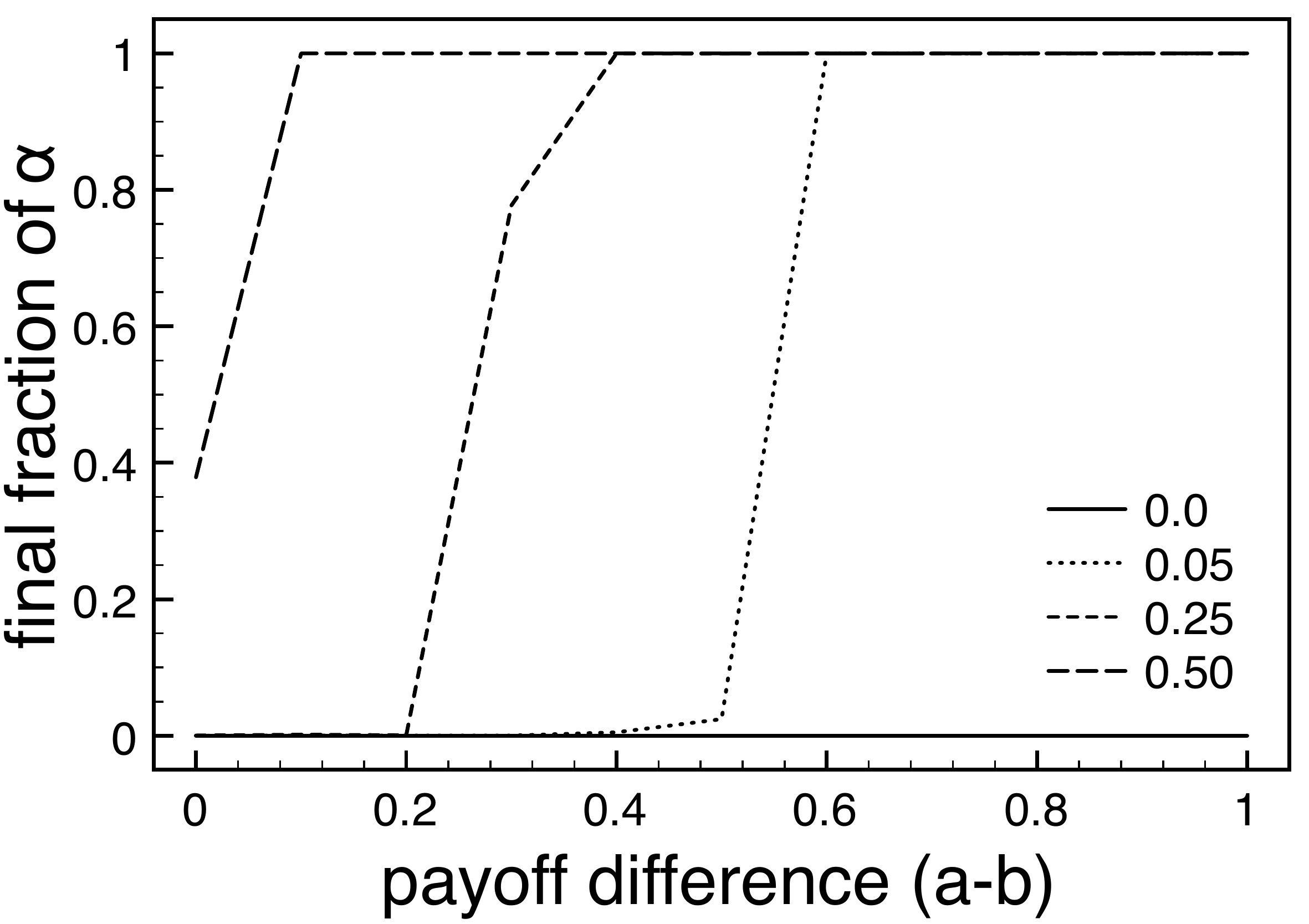} \protect &
\includegraphics[width=6.5cm] {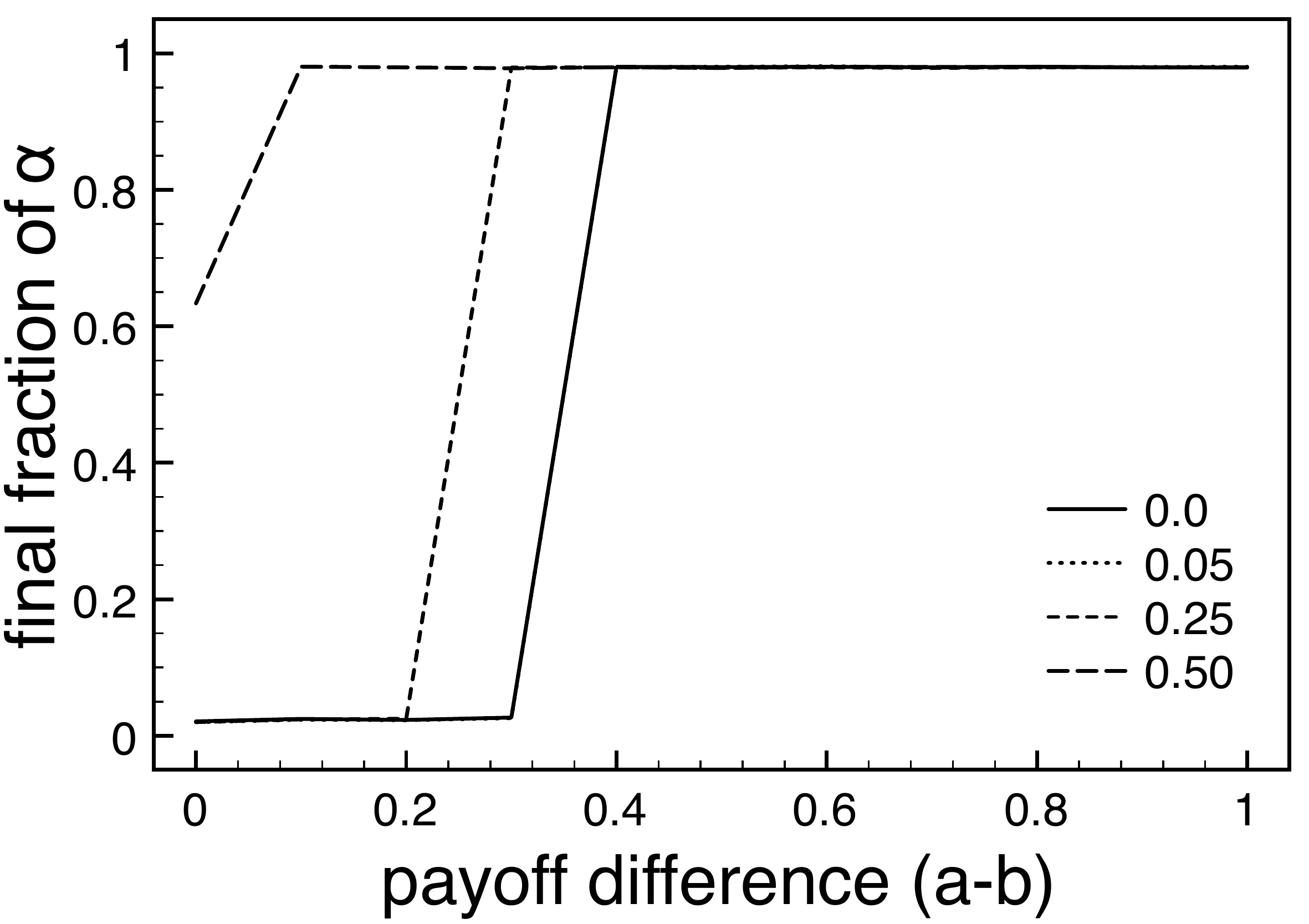} \protect \\
\end{tabular}
\caption{Random network: $\bar{k}=6$. Left image refers to noiseless best response dynamics. The right image is
for a noisy dynamics with $q=0.02$. Graphics report the frequency of strategy $\alpha$ in the population as a function
of the payoff difference $a-b$. Continuous lines are just a guide for the eye.\label{random}}
\end{center}
\end{figure}

\begin{figure} [!ht]
\begin{center}
\begin{tabular}{cc}
\includegraphics[width=6.5cm] {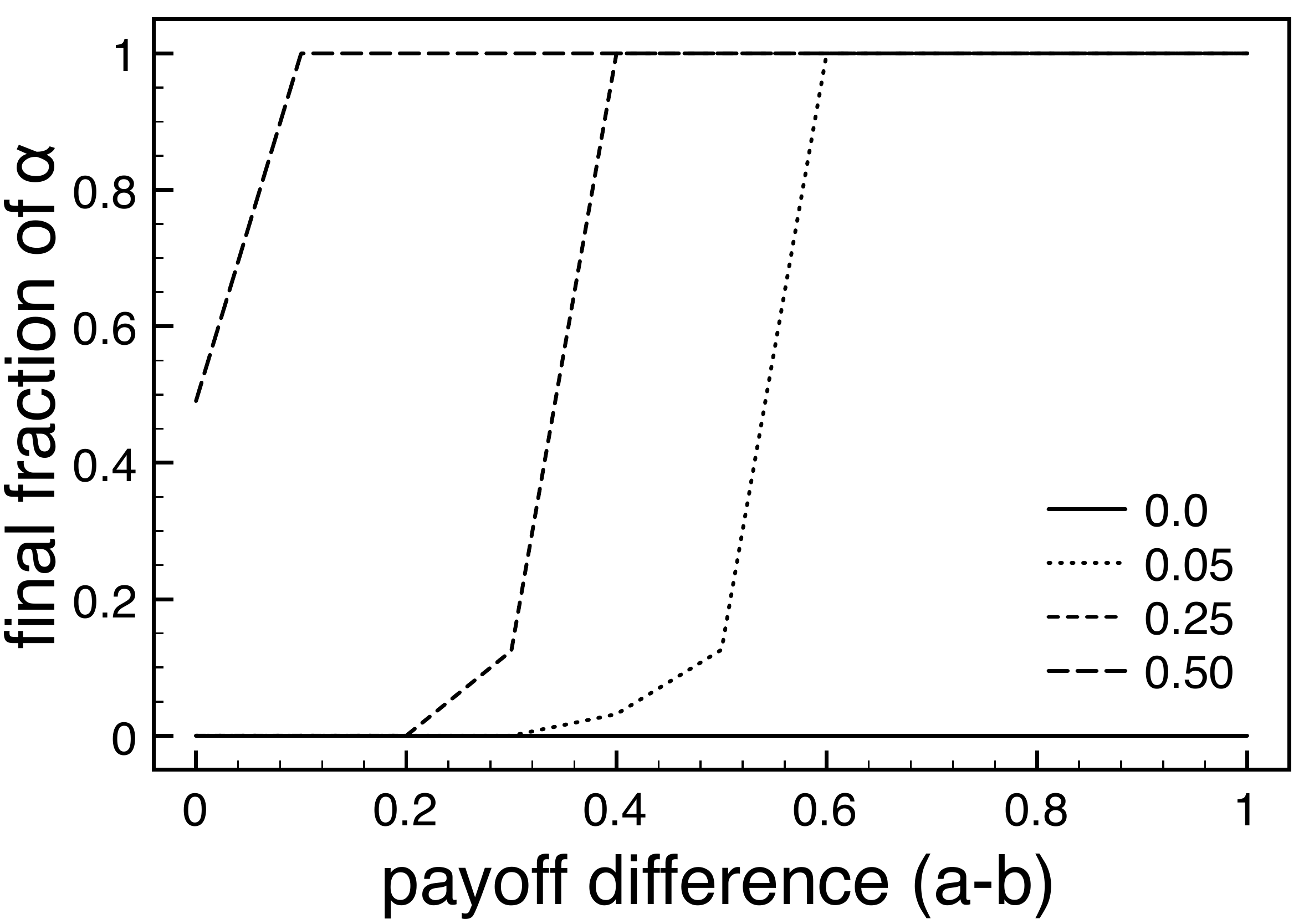} \protect &
\includegraphics[width=6.5cm] {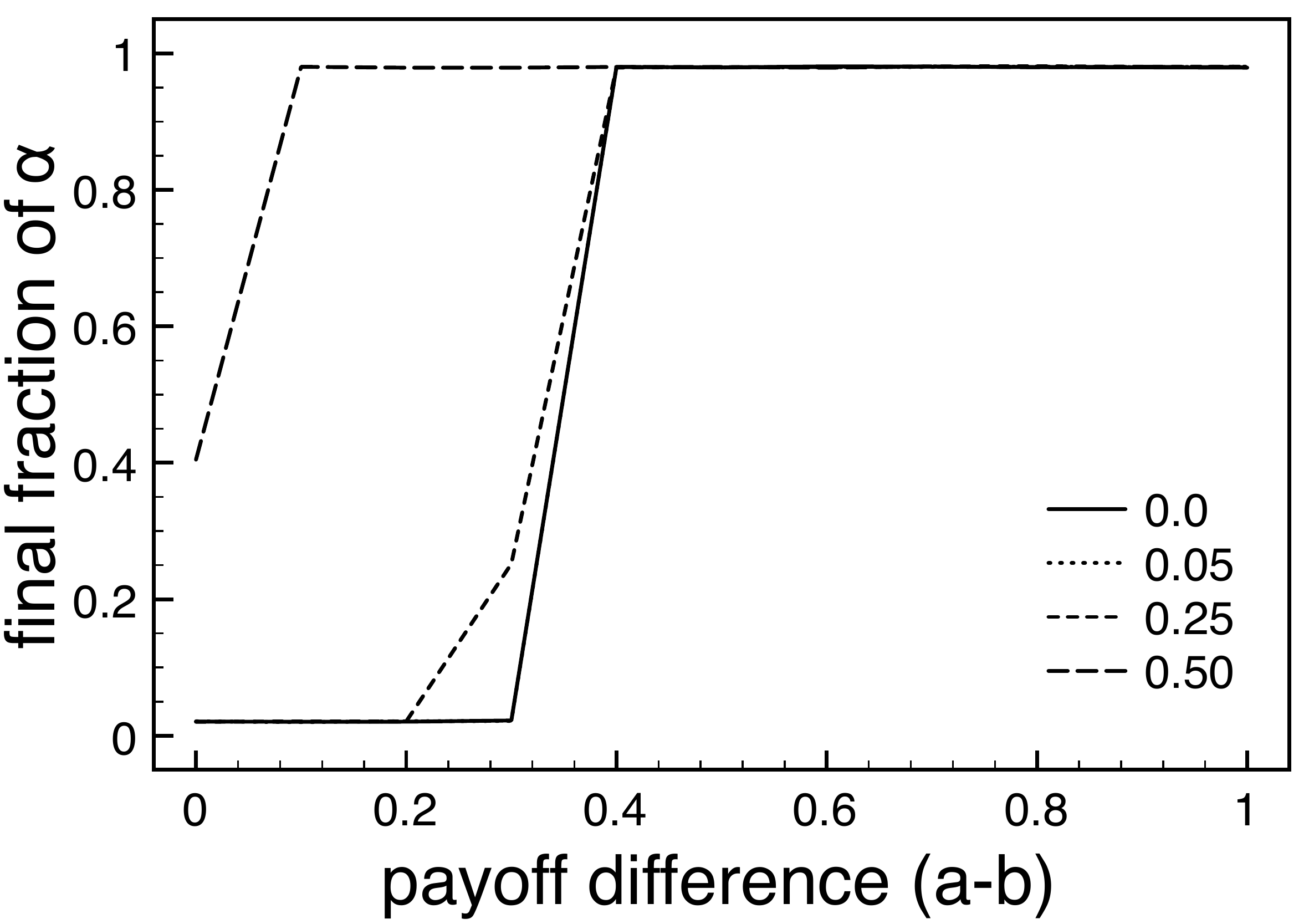} \protect \\
\end{tabular}
\caption{Scale-free network: Barabasi-Albert Model, $\bar{k}=6$. Left image refers to noiseless best response
dynamics. In the right image the probability of mutation is $q=0.02$. On the y-axis the frequency of strategy $\alpha$ is 
plotted against the payoff difference $a-b$.\label{SF}}
\end{center}
\end{figure}

Figures~\ref{random} and ~\ref{SF} show global coordination results for random graphs and scale-free
graphs respectively. The plots report on the x-axis the payoff advantage of strategy
$\alpha$ with respect to strategy $\beta$, which goes from $0$ to $1$, and  on the y-axis
the frequency of $\alpha$-strategists in the population. The curves represent average values over $50$
runs for each sampled point.
By simple inspection, it is clear that results do not differ by a large extent between the random and the scale-free cases,
which means that the degree distribution function has little effect on the outcome. The general trend is for
all the populations to converge toward the payoff-dominant Nash equilibrium in pure strategies which is also the case for the
standard well-mixed population, as we know from analytical results. Polymorphic populations do exist temporarily
but they are unstable and the stochastic dynamics always reaches a monomorphic state.
It is also quite obvious that without
mutations (Figs.~\ref{random} and~\ref{SF} left-hand images), if a strategy is absent at the beginning, it cannot appear later.
Instead, with even a small amount of noise ($q=0.02$ in the figures), the strategy offering the best payoff will take over the population
thanks to repeated mutations that will create individuals playing that strategy (Figs.~\ref{random} and~\ref{SF} right-hand images)
even in case the strategy is absent in the initial population.
Furthermore, noise always promotes a quicker transition toward the payoff-dominant steady state.

\begin{figure} [!ht]
\begin{center}
\begin{tabular}{cc}
\includegraphics[width=6.5cm] {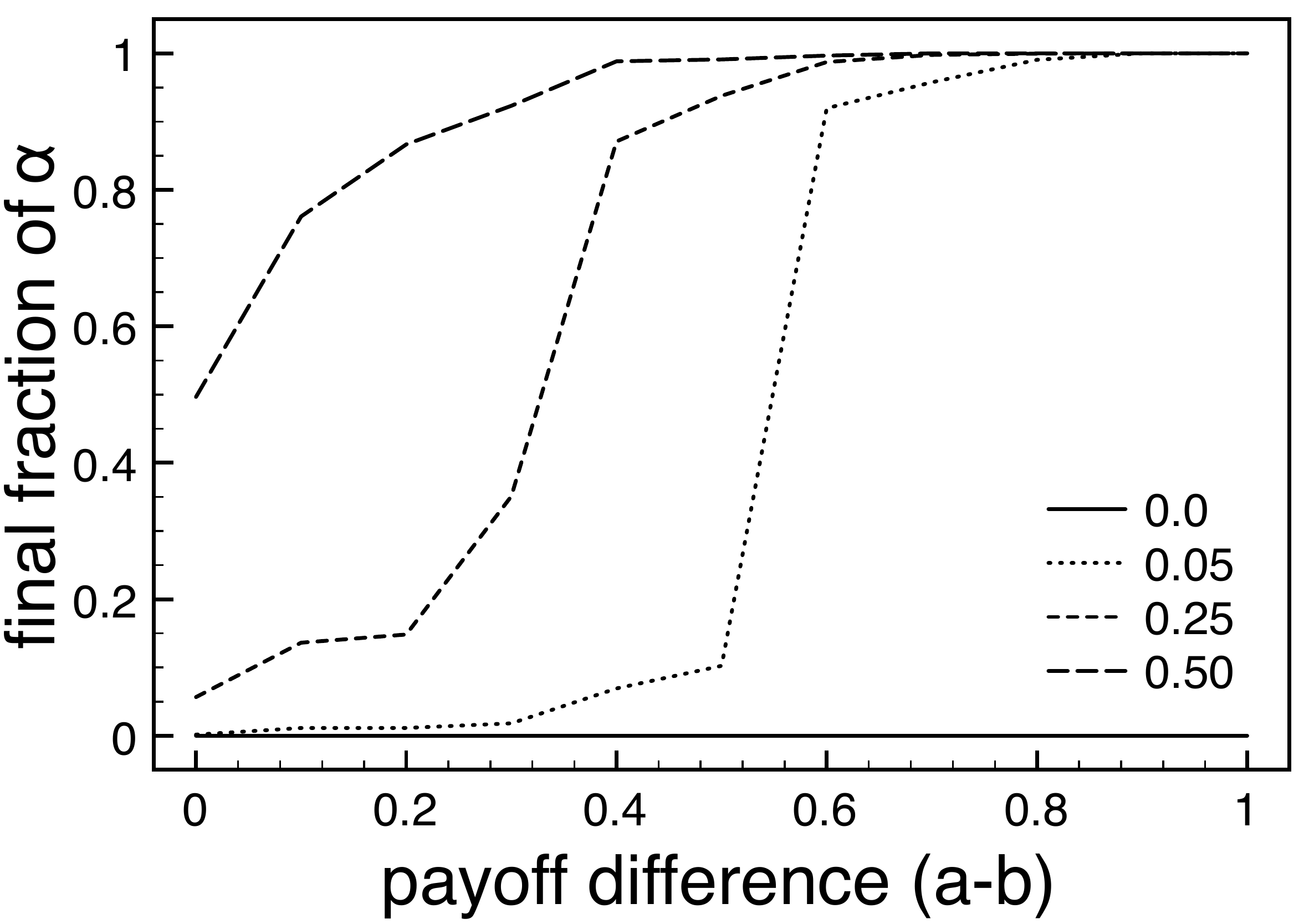} \protect &
\includegraphics[width=6.5cm] {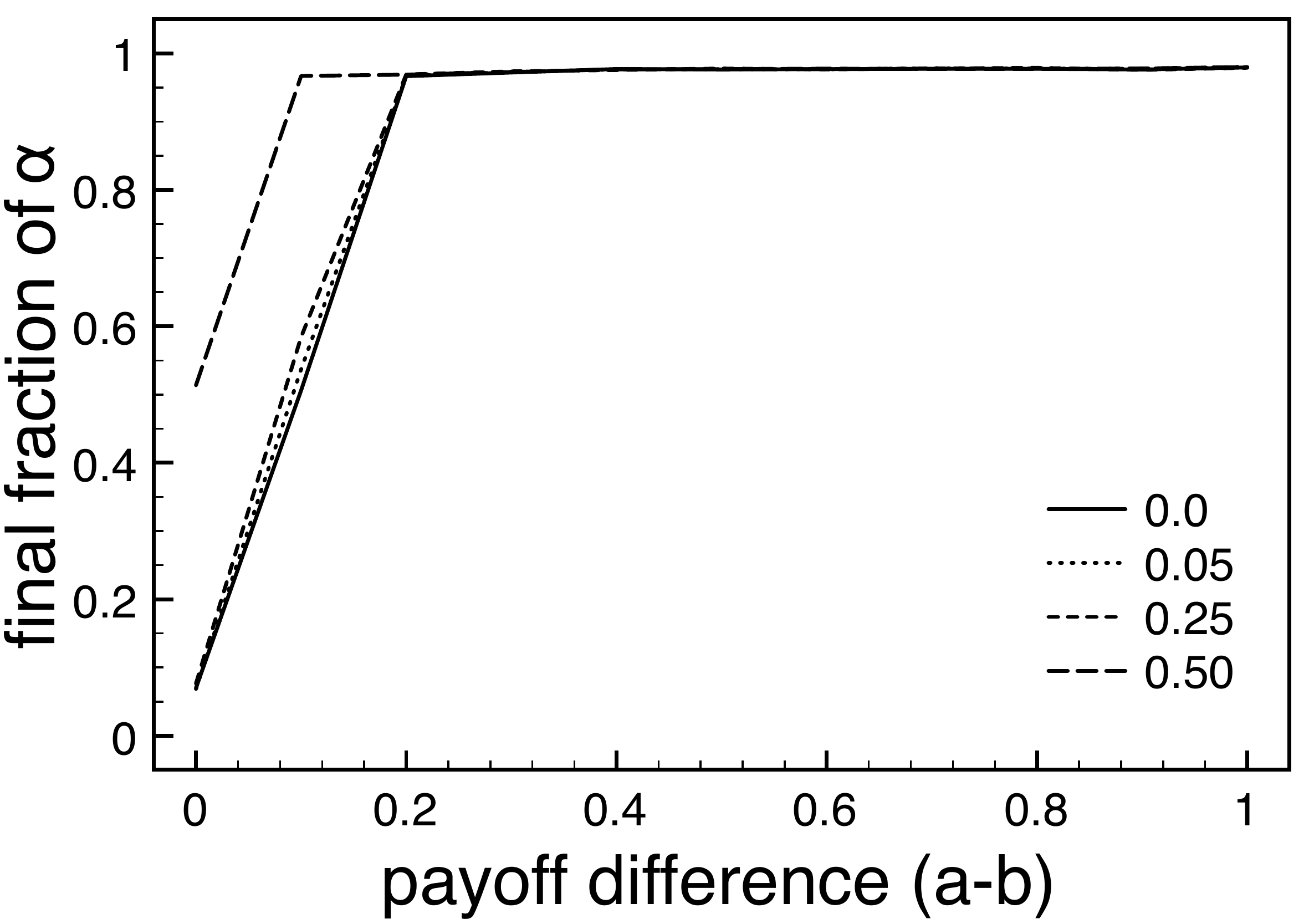} \protect \\
\end{tabular}
\caption{Coauthorship network in the Genetic Programming community. Left image: no noise.
Right image: mutation probability $q=0.02$. On the y-axis we report the fraction of $\alpha$-strategists in the population as a function of the
payoff difference $a-b$.\label{GP}}
\end{center}
\end{figure}

\begin{figure} [!ht]
\begin{center}
\begin{tabular}{cc}
\includegraphics[width=6.5cm] {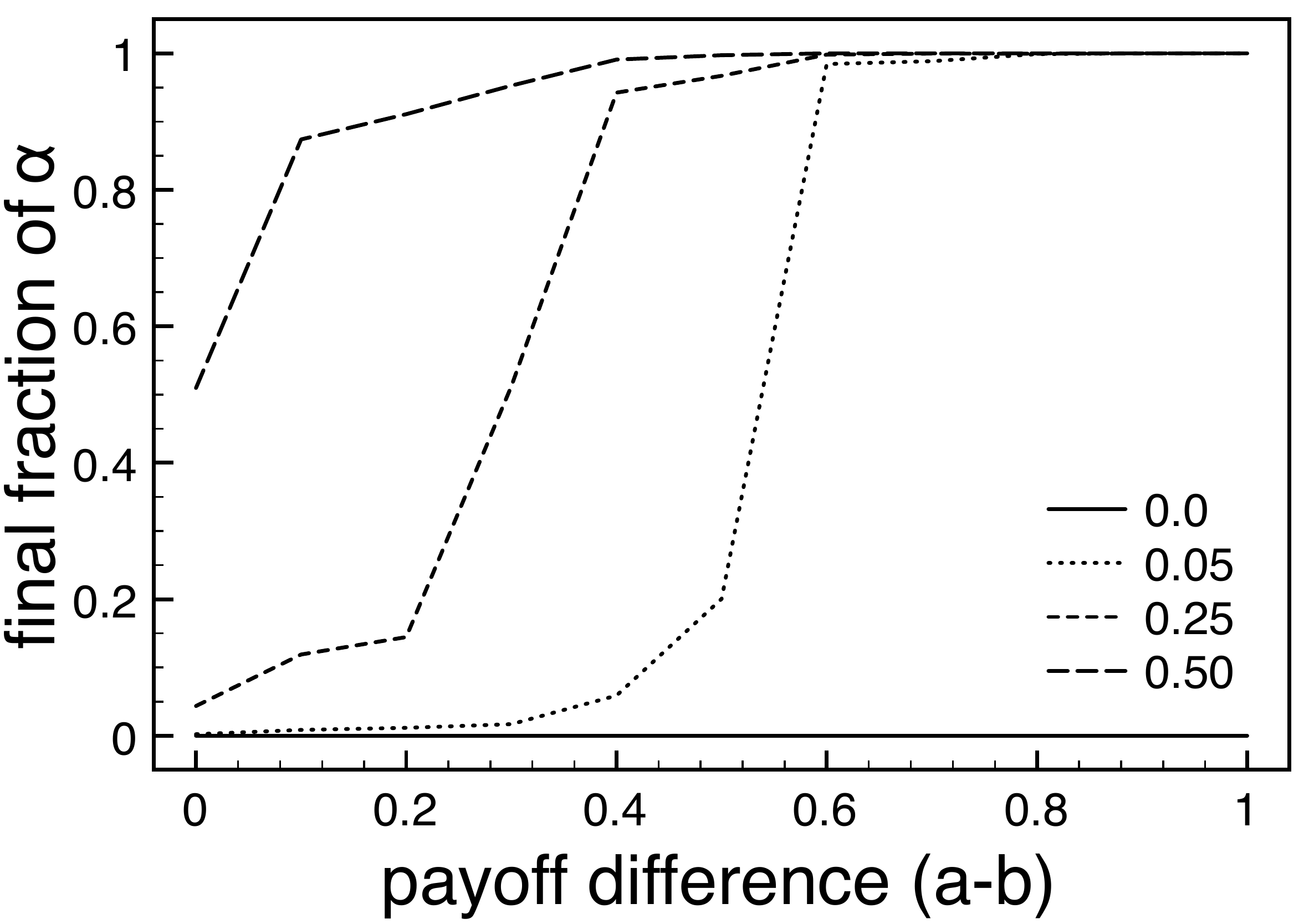} \protect &
\includegraphics[width=6.5cm] {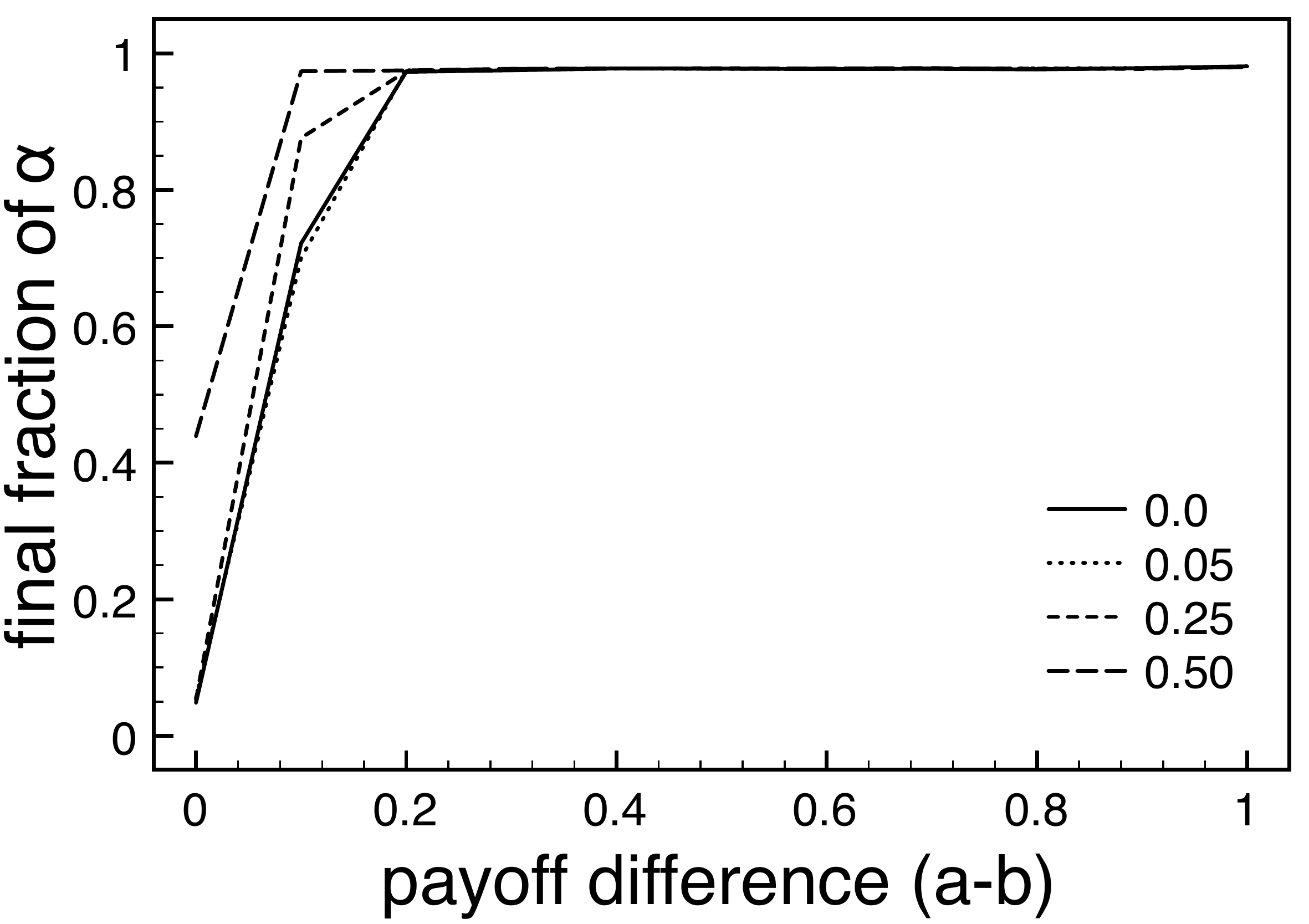} \protect \\
\end{tabular}
\caption{Model social network. Left: mutation probability $q=0$; right: $q=0.02$.
On the y-axis we report the fraction of $\alpha$-strategists in the population as a function of the
payoff difference $a-b$.\label{Toivonen}}
\end{center}
\end{figure}

Figures~\ref{GP} and~\ref{Toivonen} depict the same quantities as above in the case of the real
social network and model social networks respectively. Although the general behavior is the same,
i.e. the Pareto-dominant steady state is reached in most situations, some aspects of the dynamics
differ from the case of random and scale-free networks. To begin with, one sees on the left-hand
images that, without noise, the payoff dominated strategy is able to resist in the population when the
payoff differences are small. 
For example, starting with an equal initial share of strategies $\alpha$ and $\beta$, one sees in
Figs. ~\ref{GP} and~\ref{Toivonen} that,
up to a difference in payoffs of $0.02$ the Pareto-dominated strategy is still present in the population
with a sizable fraction. 
This phenomenon can be explained by looking at the clusters present in the social networks. Results will
be presented below.

But the main remark is that, in the presence
of noise, the payoff-dominant stable state is reached for smaller differences in payoff (see right-hand images). 
In other words,
a small $a-b$ advantage is enough to quickly steer the dynamics towards the dominant quasi-equilibrium.
The behavior is sufficiently different from the previous one to require at least a qualitative explanation, which
is presented next by introducing the concept of communities.

\subsection{Social Communities and Game Strategies}
\label{communities}
\begin{figure} [!ht]
\begin{center}
\begin{tabular}{cc}
\includegraphics[width=6.7cm] {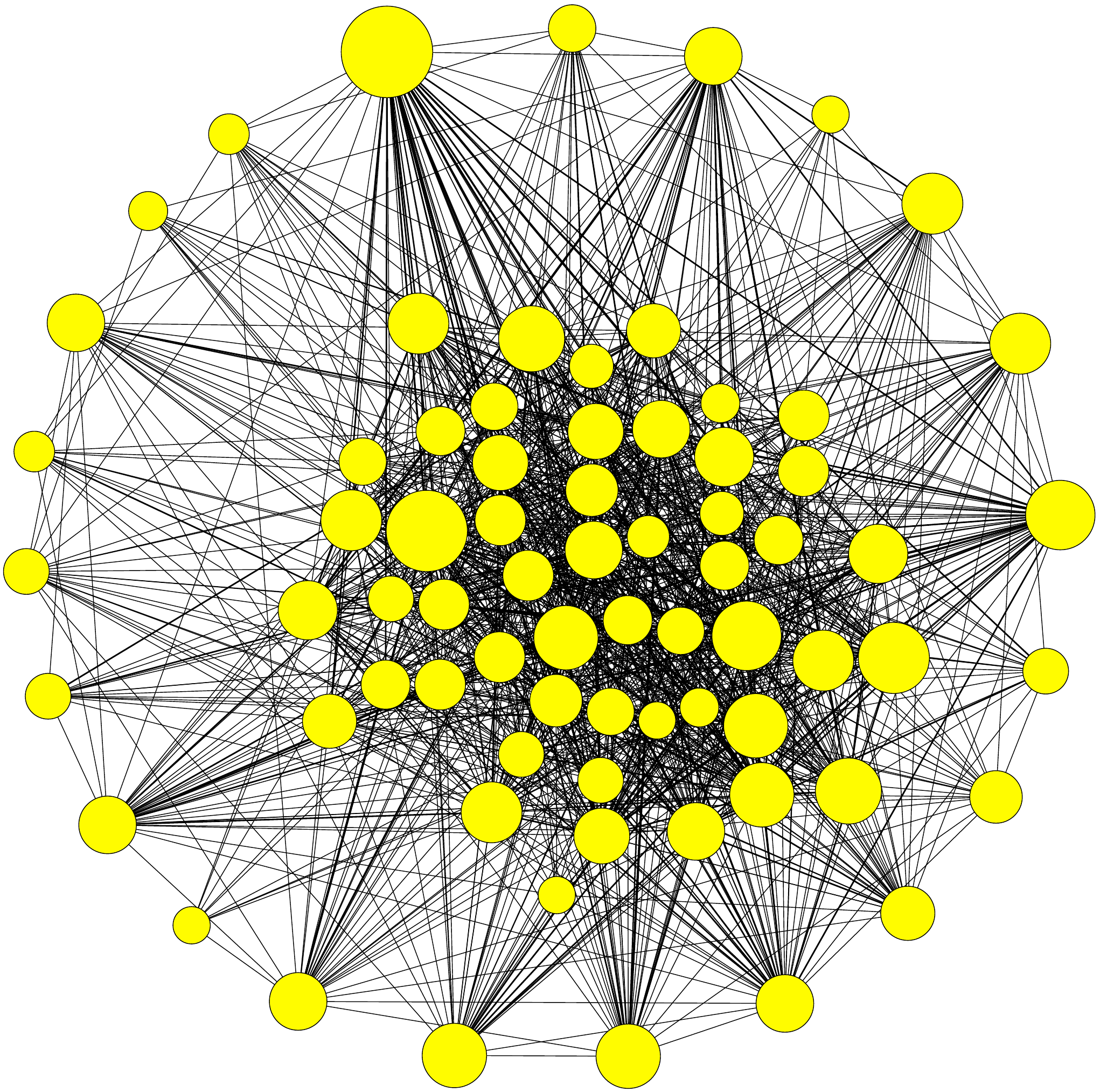} &
\includegraphics[width=6.7cm] {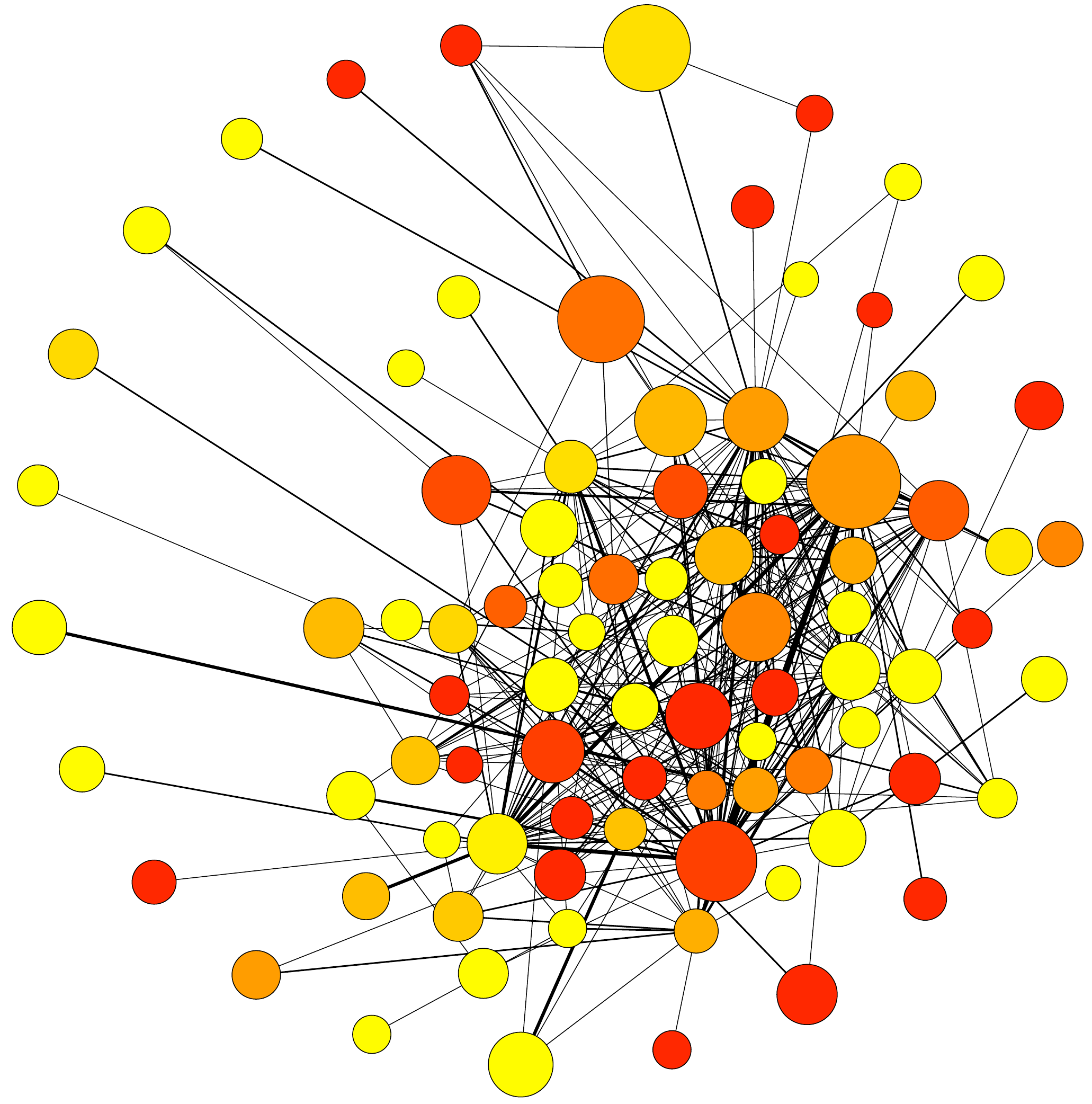} \\
(a) & (b)\\
\end{tabular}
\caption{
Distribution of  strategies at steady state in the network communities when both strategies share the same payoff: $a=b=0.5$. (a) scale-free, frequency of $\alpha=0.568$. (b) model social network, fraction of $\alpha=0.585.$ 
Each vertex
represents a whole community with size proportional to the size of the community. Links represent inter-community
connections and their thickness is proportional to the number of inter-community links. 
The communities are much less interconnected in the social network and this causes a greater difference in concentration from community to community.\label{comm1}}
\end{center}
\end{figure}
\begin{figure} [!ht]
\begin{center}
\begin{tabular}{cc}
\includegraphics[width=6.7cm] {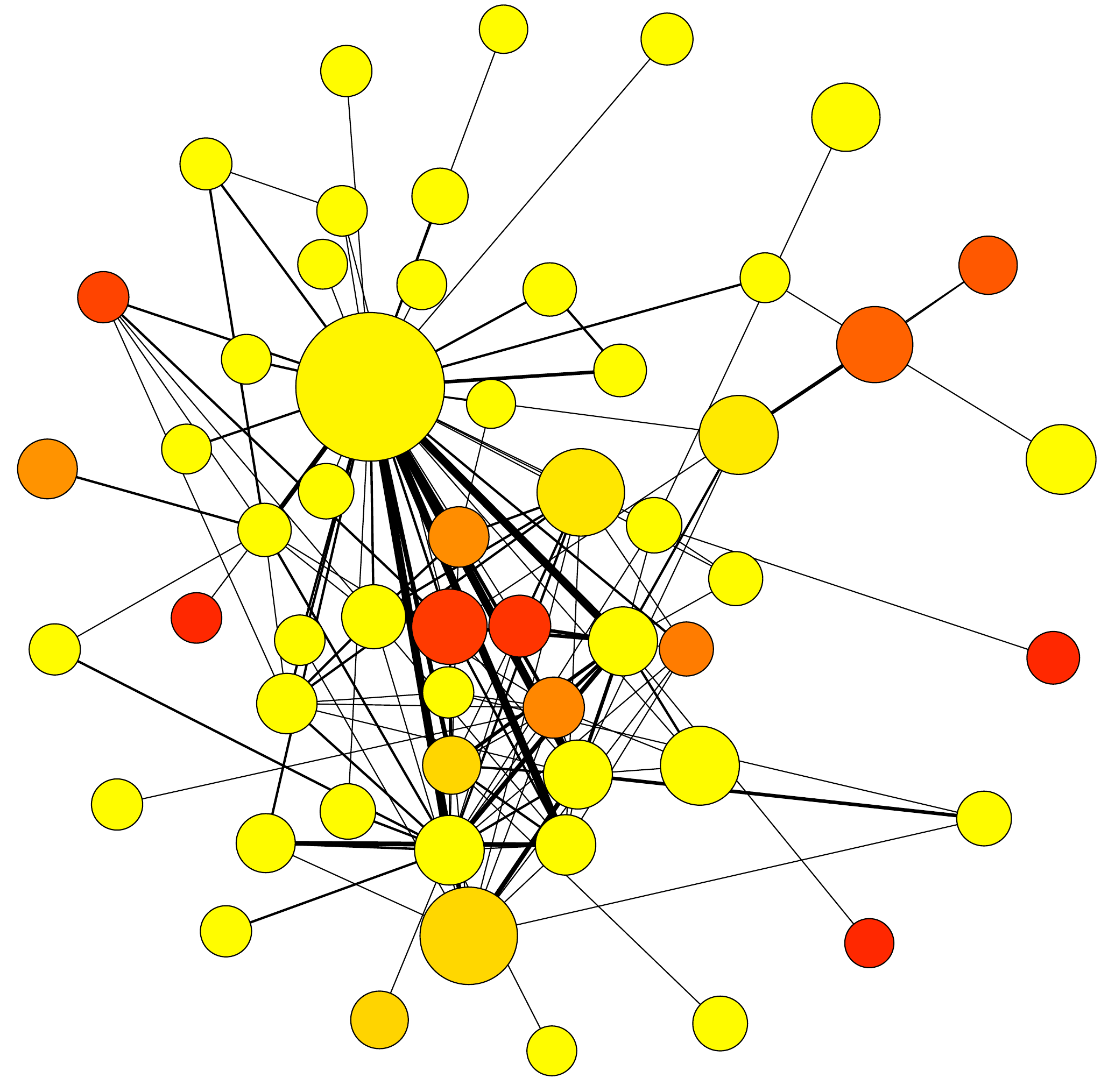} &
\includegraphics[width=6.7cm] {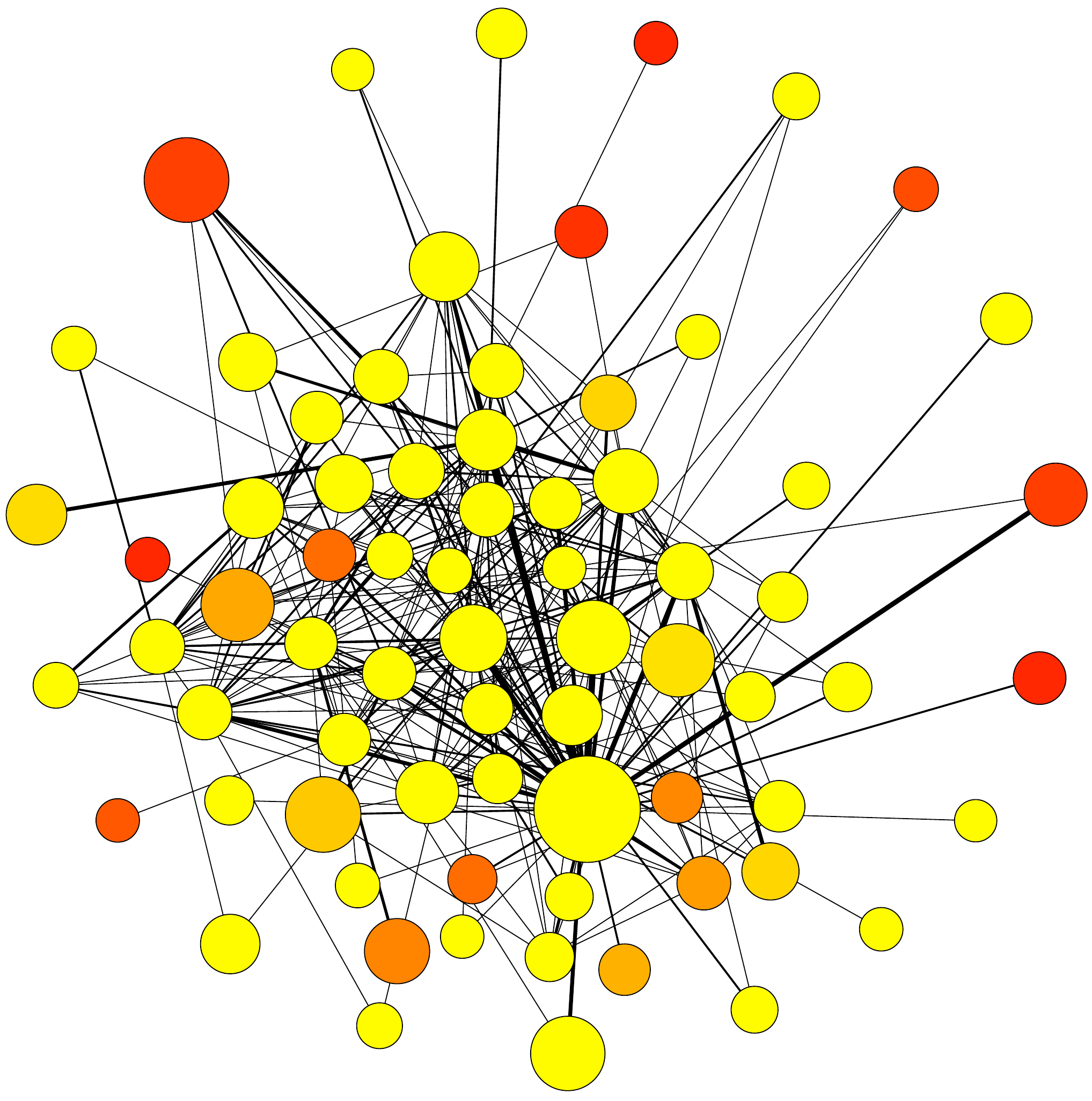} \\
(a) & (b)\\
\end{tabular}
\caption{Strategy distribution in the network communities when $\alpha$ has a small advantage over $\beta$: $a=0.55$. (a) Genetic Programming co-authorship network, proportion of $\alpha=0.839$. (b) model social network, proportion of $\alpha = 0.833$. The cluster structure of these networks allows the preservation of the dominated strategy in some communities.
\label{comm2}}
\end{center}
\end{figure}

Communities or clusters in networks can be loosely defined as being groups of nodes that are
strongly  connected between them and poorly connected with the rest of the graph. These structures are
extremely important in social networks and may determine to a large extent the properties
of dynamical processes such as diffusion, search, and rumor spreading
among others. Several methods have been proposed to uncover the clusters present in a network
(for a review see, for instance,~\cite{santo1}).
In order to study the effect of community structure on the distribution of behaviors at steady state,
 here we have used the divisive method of 
Girvan and Newman~\cite{newman-girvan-2004} which is based on iteratively removing edges with a
high value of edge betweennes. \\
The presence of communities has a marked effect on the game dynamics. Figure~\ref{comm1} depicts
the community structure of a Barab\'asi--Albert scale-free graph (a) and of a model social network built according to
Toivonen et al's model (b). The difference is striking: while clear-cut clusters exist in (b), almost no recognizable
communities can be isolated in (a), a fact that is shown by the high number of links between clusters, with
a communities graph average degree of $\sim32$, while $\bar k$ is about $6.5$ for the communities graphs
arising from social networks. 
A common statistical indicator of the presence of a recognizable community structure is the \textit{modularity} $Q$.
According to Newman~\cite{newman-06}, where quantitative definitions are given, 
modularity is proportional to the number of edges falling within clusters minus the expected number in an equivalent network 
with edges placed at random. While modularity is not without flaws~\cite{guimera}, it is still a convenient indicator of the presence of clusters.
In general, networks with strong community structure tend to have values of $Q$ in the range $0.4-0.7$. Indeed, for the
networks in Fig.~\ref{comm1}, we have $Q \simeq 0.3$ for the scale-free network, while $Q \simeq 0.6$ for the model social
network.
Colors in the
figure represent frequency of strategies at steady state for a single particular, but representative, run in each case.
In the average over $50$ runs, final proportions of strategies $\alpha$ and $\beta$ do not depart much from the initial $50\%$.
However, while in the scale-free case at the steady state the standard deviation is high, meaning that the
system converges often to one or the other equilibrium, this is not the case for the social networks. In the latter,
at steady state there is always a mix of strategies; in other words, polymorphic equilibria may be stable.
This is a remarkable fact that is due to the community structure of social networks, which is almost missing in the scale-free 
and random network cases. Thanks to this clear-cut cluster structure, as soon as the nodes of a cluster are colonized by a 
majority of one of the two strategies  by statistical fluctuation, it becomes difficult for the other strategy to overtake, which
explains why these cluster strategies are robust. The effect of the community structure is even more apparent in
Fig.~\ref{comm2} where strategy $\alpha$ has been given a slight initial advantage. At steady state,
in both the co-authorship network (a) as well as the model network (b) strategy $\beta$ is still present in some clusters.
If we were to interpret strategies as social norms or conventions, then this would suggest that a realistic social
structure may help protect diversity, either political or cultural, for example. The possibility of polymorphic equilibria had
been theoretically predicted by Morris~\cite{morris} for symmetric payoffs in pure coordination games with best
response dynamics in the case of infinite populations and making use of a notion of ``cohesion'' which refers to
the relative frequency of ties among groups compared with non-members. Clearly, although it was expressed in
a different language that does not make explicit use of networks, this notion is related to the communities we
have here and the simulation results nicely confirm the prediction in the case of finite, actual networked systems.

\subsection{Results on the Stag Hunt Games}
\label{res-stag-hunt}

\begin{figure} [!ht]
\begin{center}
\includegraphics[width=12cm] {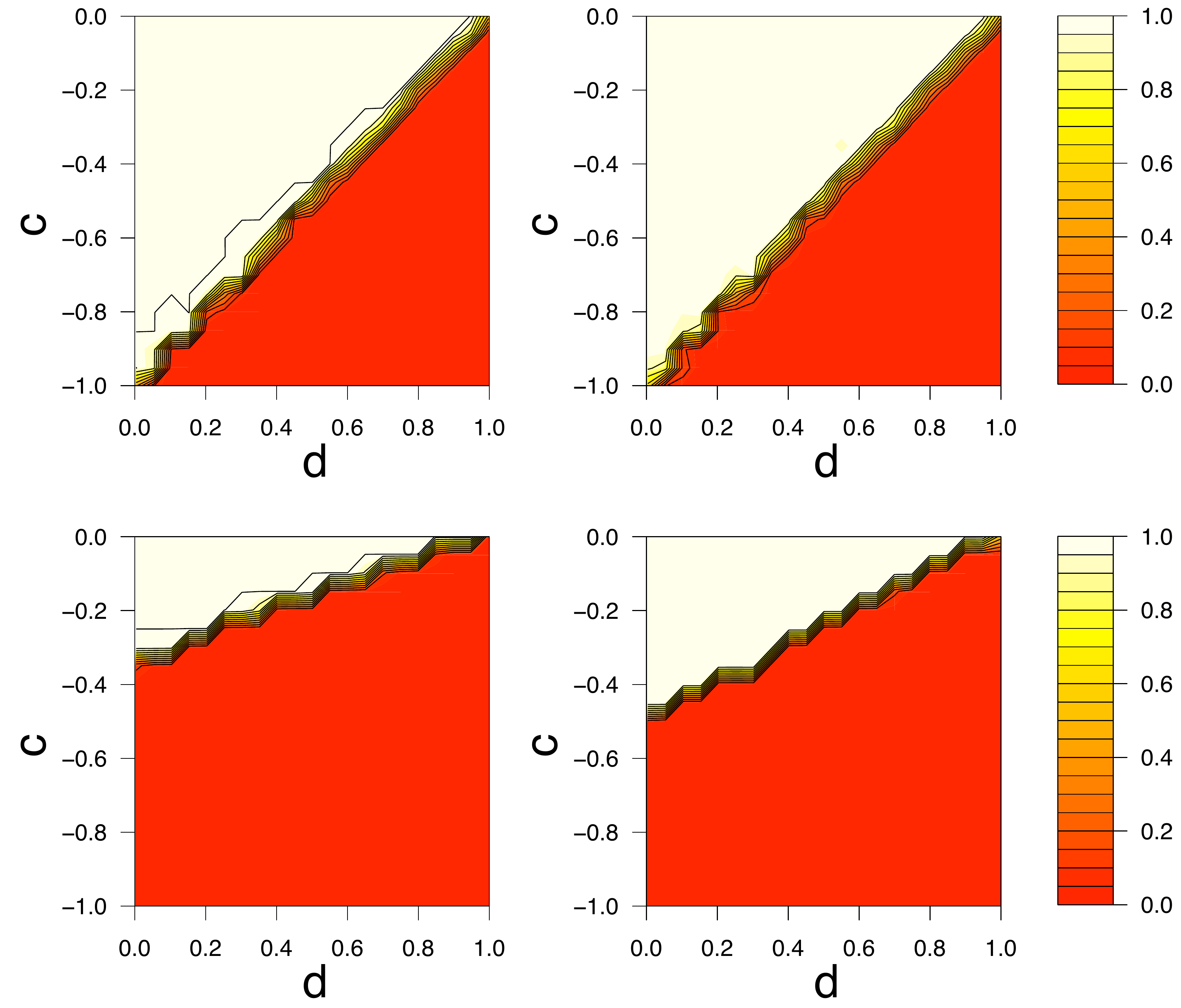}
\caption{Distribution of strategies proportions at steady state on the $d-c$ plane for scale-free networks. Each sampled point is
the average of $50$ independent runs. The upper images are for equal initial proportions of each strategy. In the lower figures
the initial proportion of randomly distributed $\alpha$-strategists is $5\%$. Figures on the left column are for best response
dynamics without noise, while those on the right column represent a situation in which the probability of mutation $q=0.02$.
Darker colors mean that risk-dominance prevails; light color design the
region where payoff-dominance prevails. \label{sh}}
\end{center}
\end{figure}

\begin{figure} [!ht]
\begin{center}
\includegraphics[width=12cm] {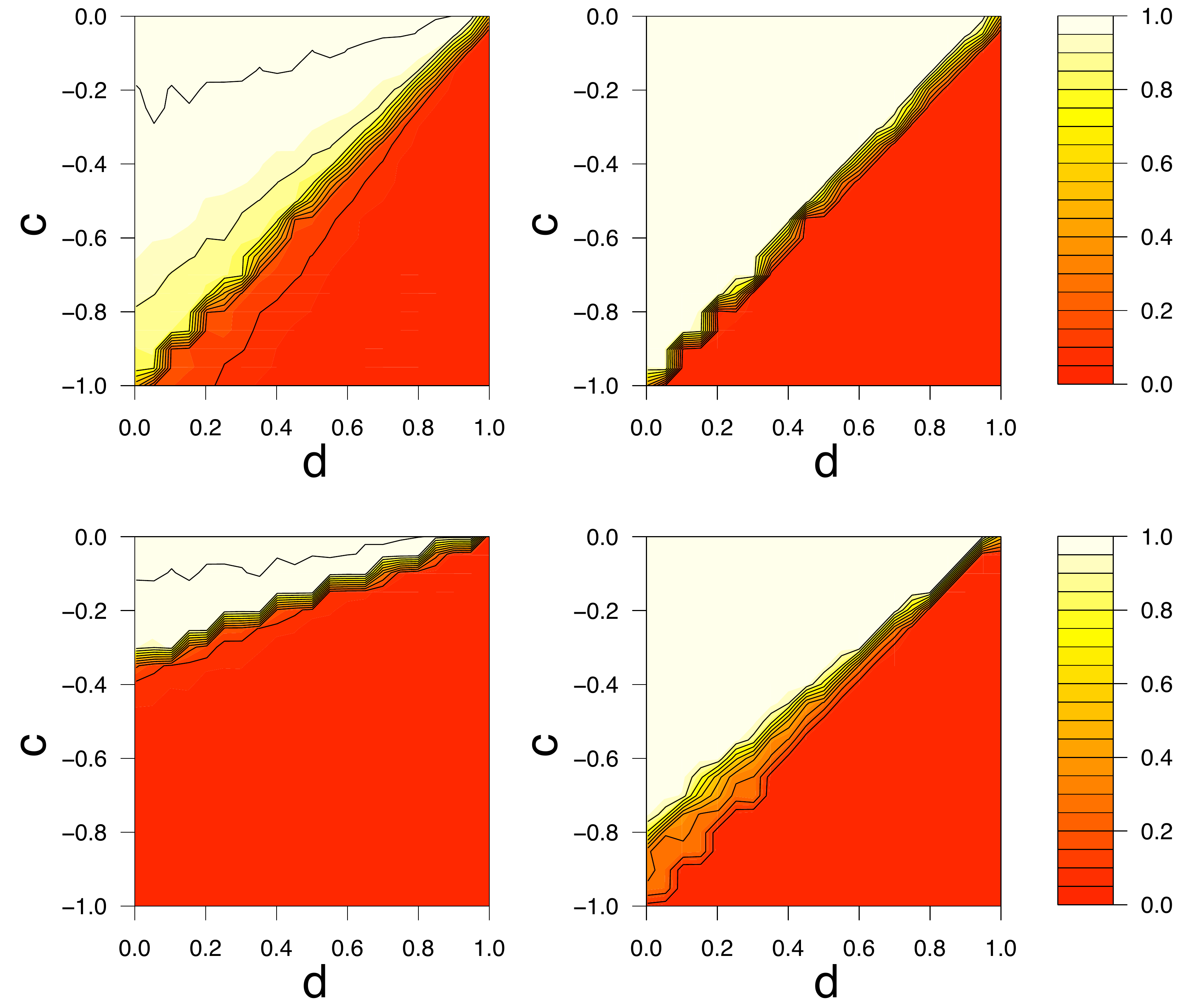}
\caption{Distribution of strategies proportions at steady state on the $d-c$ plane for model social networks. Each sampled point is
the average of $50$ independent runs. The upper images are for equal initial proportions of each strategy. In the lower figures
the initial proportion of randomly distributed $\alpha$-strategists is $5\%$. Figures on the left column are for best response
dynamics without noise, while those on the right column represent a situation in which the probability of mutation $q=0.02$.
Darker colors mean that risk-dominance prevails; light color design the
region where payoff-dominance prevails.\label{sh2}}
\end{center}
\end{figure}

Figure~\ref{sh} shows strategy distribution on the game parameter space for the Stag Hunt class of
coordination games for the scale-free case. Results
for random graphs are similar to those for scale-free networks and are not shown.
The two upper images are for equal initial proportions of each strategy, while the bottom figures
refer to an initial proportion of strategy $\alpha=5\%$. The first image in each row is for best response without
noise, while the second image 
has noise level $q=0.02$. \\
For initially equidistributed strategies, although average values are reported in the figures, almost all simulations attain one
or the other absorbing state, i.e. all individuals play $\alpha$ or all play $\beta$, and there is almost no difference when noise is present. This is in agreement
with previous results on scale-free graphs published by Roca et al.~\cite{anxo2} where update was by best response without noise,
and also with~\cite{luthi-pest-tom-physa08} where replicator dynamics instead of best response dynamics
was used as a strategy update rule.\\
For the more extreme case in which initially the fraction of 
strategy $\alpha$ is $5\%$ randomly distributed over the graph vertices (bottom row images),
a small amount of random noise does not have a large effect: the cooperative strategy emerges in the favorable
region of the parameter space, i.e.  for low $d$ and high $c$ (upper left corner) in both cases.
However, the presence of noise enhances the efficient coordination region. Indeed, even when strategy
$\alpha$ is initially absent, once it is created by mutation, it spreads as in the $5\%$ case.
 It is to be noted that the same
phenomenon happens when the minority strategy is $\beta=0.05$; in this case the images are specularly symmetrical,
and with colors reversed,
with respect to the main diagonal, except for sampling differences (not shown to save space).\\
Figure~\ref{sh2} depicts average results for the model social network case of 
Toivonen et al.~\cite{toivonen-2006}.  Results for the collaboration network are very close to those of model social networks. 
For this reason, and in order not to clutter the graphics too much, we do not show them. 
It is immediately apparent that the case in which strategies are initially randomly distributed in equal
amounts seems similar to the scale-free results. However, looking more carefully, the average results shown in the figures
hide to some extent the fact that now many simulations do not end in one of the monomorphic population
states, but rather there is a mix of the two strategies, when noise is absent. This is visible in the upper left figure in the less crisp frontier
along the diagonal which is due to a more gradual transition between phase space regions. However, when
a small amount of noise is present (upper right image) the transition is again sharp and the dynamics usually leads to
a monomorphic population in which one of the two absorbing states is entered. The reason why 
there can be mixed states in the noiseless case in social networks is related to their mesoscopic structure. 
As we have seen in sect.~\ref{communities}, model and real social networks can be partitioned
into recognizable clusters. Within these clusters strategies may become dominant as in the pure
coordination case just by chance. In other words, as soon as a strategy dominates in a given cluster,
it is difficult to eradicate it from outside since other communities, being weakly connected, have little
influence. 
This kind of effect in the Stag Hunt game has been observed previously 
in simulations on grid-structured populations~\cite{skyrms,anxo2}. However, grid structures are not socially realistic; thus,
the fact that more likely social structure do support efficient outcomes is an encouraging result. 
However, when noise is present, there is always the possibility that the other strategy appears in the cluster by
statistical fluctuations and, from there, it can takeover the whole community.
To end this section, we remark that analogous effects due to the presence of clusters in social networks
have been observed and interpreted in the  Prisoner's Dilemma game 
in~\cite{arenas-comm-08,luthi-pest-tom-physa08}.

We now briefly comment on the relationship between our numerical results and well known theoretical 
results on Stag-Hunt games. These
theoretical models are based on ergodic stochastic processes in large populations and state that, when using
best-response dynamics in random two-person encounters, and in the presence of a little amount
of noise, both for well-mixed
populations as well as for populations structured as rings, the risk-dominant strategy should take
over the population in the long run~\cite{kandori,ellison,young}. From our simulation results on all kind
of networks this is not the case; in other words, at the steady state there is always either a single strategy,
but not necessarily the risk-dominant one, or a mix of both strategies. For scale-free and random
graphs, the numerical results of~\cite{anxo2} agree with ours. The case of social networks, presented
here for the first time, also confirms the above and in addition makes explicit the role played by communities. 
We may also mention at this point that, for the Stag-Hunt, the presence of a local interaction structure
provided by a network has been shown to
increase the region of the phase space in which the Pareto-dominant outcome prevails for other
strategy update rules, such as imitate the most successful neighbor or reproduce proportionally to
fitness (replicator dynamics)~\cite{skyrms,anxo1}. Thus coordination is sensitive to the exact type
of underlying dynamics in networks. This is indirectly confirmed by the theoretical study of Robson and 
Vega-Redondo~\cite{robson-vega} in which a different matching model is used with respect to Kandori et al~\cite{kandori}.
In~\cite{robson-vega} players are immediately
randomly rematched after each encounter and the result is that the Pareto-dominant equilibrium is selected instead.\\
In summary, it can be said that network effects tend to reinforce cooperation on the Pareto-dominant
case, which is a socially appreciable effect. However, these results must be taken with a grain of salt.
We are numerically studying finite, network-structured populations during a limited amount of time, while 
theoretical results have
been established for large well mixed populations in the very long run. The conditions are thus sufficiently
different to conclude that numerical results  and theoretical predictions based on different 
assumptions do not have to agree necessarily.

\section{Summary and Conclusions}
\label{concl}

In this work we have studied pure and general coordination games on complex networks by numerical simulation. Situations described
by coordination games are common in society and it is important to understand when and how coordination on
socially efficient outcomes can be achieved. \\
In the case of pure coordination games on model networks using deterministic best response strategy
dynamics we have found that network effects are small or non-existent in standard complex networks.
On model social networks and a real co-authorship network the behavior is similar, but the transition from
one convention to the other is smoother and the cluster structure of the networks plays an important role
in protecting payoff-weaker conventions within communities and this leads to a clear polarization of conventions in the network.
When a small amount of noise is added in order to simulate 
errors and trembles in the agent's decisions, the dynamics leads to the payoff-dominant norm for smaller
values of the payoff difference between strategies.
However, in the case of social networks, even a tiny amount of payoff advantage is enough to drive 
a minority of $\alpha$-strategists to take over the whole network thanks to the cluster structure and mutations.\\
In the case of general coordination games of the Stug Hunt type where there is a tension between payoff-dominance
and risk-dominance, we have confirmed previous simulation results in the sense that, with deterministic best
response dynamics the influence of network structure is very limited~\cite{skyrms,anxo2,anxo1,luthi-pest-tom-physa08}. 
On the other hand, when we consider model
and social networks, again their community structure plays an important role which consists in allowing the
existence at steady state of dimorphic populations in which both strategies are present and stable.  
The payoff-dominant strategy is favored in regions where risk-dominance should be the only stable strategy and,
conversely, it allows risk-dominant players to survive in clusters when payoff-dominance should prevail.\\
We have also compared numerical results with theoretical ones when they exist. The latter actually depend
on the detailed structure of the stochastic processes generated by the particular theoretical model. In this sense,
numerical results are compatible with theoretical predictions when they are applicable, i.e. for well mixed and
ring-structured populations~\cite{kandori,robson-vega,ellison}. Also, for pure coordination games the predictions
of~\cite{morris} in arbitrary non-homogeneous structures are qualitatively confirmed. However, finite-size and 
complex network effects are difficult to describe theoretically 
and thus our results on complex and social networks cannot always be easily compared with theoretical
predictions. Our current and future work is to investigate coordination games in a more realistic co-evolutionary scenario in
which both the agents' strategies as well as their interactions may vary dynamically.

\paragraph*{Acknowledgments.} We thank Rafael Lalive for stimulating discussions and for reading the manuscript. 
We gratefully acknowledge financial support by the Swiss National Science Foundation under contract 200020-119719.

\end{document}